# Gate-tunable superconducting quantum interference devices of PbS nanowires

Hong-Seok Kim[1], Bum-Kyu Kim[1], Yiming Yang[2], Xingyue Peng[2], Soon-Gul Lee[1], Dong Yu[2], Yong-Joo Doh[1,*]

[1]*Department of Applied Physics, Korea University Sejong Campus, Sejong, 339-700, Republic of Korea*
[2]*Department of Physics, University of California, Davis, CA 95616, USA*

[*]E-mail: yjdoh@korea.ac.kr

**Abstract**

We report on the fabrication and electrical transport properties of gate-tunable superconducting quantum interference devices (SQUIDs), made of semiconducting PbS nanowire contacted with PbIn superconducting electrodes. Applied with a magnetic field perpendicular to the plane of the nano-hybrid SQUID, periodic oscillations of the critical current due to the flux quantization in SQUID are observed up to $T$ = 4.0 K. Nonsinusoidal current-phase relationship is obtained as a function of temperature and gate voltage, which is consistent with a short and diffusive junction model.

Nano-hybrid superconducting junctions,[1-2] made of a single-crystalline nanostructure contacted with conventional superconductors, provide a useful tool to explore the combined effects of quantum electrical transport in nanostructure and superconducting phase coherence. Supercurrent transistors, in which the superconducting coupling through the nanostructure is gate-tunable, have been developed using semiconductor nanowires (NWs),[1, 3] carbon nanotube,[4] and graphene[5]. Controlled Cooper pair splitters for nonlocal entanglement have also been realized using the nanostructures.[6-7] Furthermore, recent studies on the gate-tunable macroscopic quantum tunneling[8-9] and Majorana bound states[10-11] in the nano-hybrid superconducting junctions would pave the way for developing nano-hybrid superconducting qubits.[12-13]

A superconducting quantum interference device (SQUID),[14] formed with two superconducting junctions connected in parallel, is used as a very sensitive magnetometer and a key building block for a superconducting flux qubit[15] as well. So far, nano-hybrid SQUIDs have been made of various nanostructures such as NWs,[16] carbon nanotube,[17] and graphene.[18] Very low operation temperatures below $T$ = 2.5 K,[19] however, hinders their wide applications in superconducting electronics and quantum information devices. In addition, gate-tunable current-phase relation (CPR) has not yet been studied in the NW-based superconducting junctions.[20] CPR in the superconducting weak links is expected to be nonsinusoidal,[21] while tunneling-type superconducting junctions exhibit a sinusoidal one, $I_s = I_c\sin\phi$, where $I_s$ is the supercurrent, $I_c$ is the critical current, and $\phi$ is the phase difference between two superconducting electrodes.[22] Since the CPR determines the shape of an anharmonic potential well for the Josephson phase particle,[23] CPR in the NW-based superconducting junctions would be important for developing gate-tunable superconducting qubits.

In this letter, we report on the gate-tunable operation of nano-hybrid SQUIDs, made of PbS NW and PbIn superconductor. Employing PbIn alloy as superconducting electrodes[24-25] enables us to achieve higher operation temperature of the NW SQUIDs above the liquid-helium temperature, which is the highest reported to date. Modulation of $I_c$ as a function of magnetic flux through the SQUID loop was obtained with varying temperature and gate voltage, resulting in a nonsinusoidal CPR at lower temperatures. Our observations are consistent with a short and diffusive junction model and suggest that the skewness of the CPR can be controlled by the application of gate voltage.

PbS NWs were synthesized via a chemical-vapor-deposition method in a tube furnace, as described elsewhere[26] (also see Supplementary Data). NW-based SQUIDs with different loop areas were fabricated by electron-beam lithography using PbS NWs (see Fig. S1), as the details are explained in Supplementary Data. Figure 1a shows a scanning electron microscopy (SEM) image of a typical NW SQUID, which consists of two superconducting junctions along the NW and a supercurrent loop. Electrical transport properties of NW SQUID are measured by a four-point measurement in a closed-cycle helium cryostat and $^3$He refrigerator (Cryogenic Ltd.) down to the base temperatures of 2.6 K and 0.3 K, respectively.

The PbIn electrodes become fully superconducting below $T_c$ = 6.7 K (see Fig. S1), and the supercurrent through the PbS NW SQUID is observable up to $T$ = 5.2 K, as shown in the inset of Figure 1b. The maximum operation temperature in this work is two times higher than that of InAs NW SQUID,[19] which is attributed to a very strong Josephson coupling between PbS NW and superconducting PbIn electrodes.[27] When the magnetic field is applied perpendicular to the NW SQUID loop, gradual changes in the current-voltage (*I-V*) characteristic curves are shown in Fig. 1b with increasing magnetic flux $\Phi$. It is clearly shown that $I_c$ is maximum at $\Phi = 0$ and absent at $\Phi_0/2$, where $\Phi_0 = h/2e$ is the magnetic flux quantum, $h$ is Planck's constant and $e$ is the elementary charge. For $\Phi > \Phi_0/2$, $I_c$ increases to reach its maximum value at $\Phi = \Phi_0$ and then exhibits an oscillatory behavior with a period of $\Phi_0$ as a signature of the SQUID. The periodic modulation of $I_c$ is displayed in a color plot of differential resistance, d$V$/d$I$, in Fig. 1c. Here we used the effective area of the SQUID loop to be $A = 3.39$ $\mu$m$^2$ (the yellow dashed line in Figure 1a), which is calculated from the magnetic-field periodicity of $H_0 = 6.1$ Oe. Difference between the effective area and the geometrical inner area of the loop is due to the London penetration depth, $\lambda$, of PbIn electrodes,[24] which is estimated to be $\lambda = 0.43$ μm for D1 and 0.48 μm for D2.

The periodic modulation of $I_c(\Phi)$ can be explained by the sinusoidal current-phase relation (CPR) in the NW Josephson junction and flux quantization in the SQUID loop.[22] Under the assumption of negligible self-inductance of the loop, $I_c$ is given by

$$I_c(\Phi) = [(I_{c1} - I_{c2})^2 + 4I_{c1}I_{c2}\cos^2(\pi\Phi/\Phi_0)]^{1/2} \qquad (1),$$

where $I_{c1}$ and $I_{c2}$ are the critical currents of each weak link.[19] At $T = 2.8$ K, $I_{c1} = 298$ nA and $I_{c2} = 295$ nA are obtained by fitting Eq.1 to experimental data (see white line in Fig. 1c). It is inferred that two weak links in the NW SQUID are almost identical with $I_{c2}/I_{c1} = 0.99$. The self-inductance is calculated to be $L_S \sim 3$ pH from the SQUID geometry, corresponding to the screening parameter $\beta_L = 2\pi L_S I_0/\Phi_0 = 3 \times 10^{-3} \ll 1$. Here, $I_0 = (I_{c1} + I_{c2})/2$ is an average critical current. We note that the periodic $I_c(\Phi)$ relation is maintained up to $T = 4.0$ K, as shown in Fig. 1d, which confirms the sinusoidal CPR in the NW weak links at higher temperatures.

The NW SQUID, as a flux-to-voltage transducer, can provide an output voltage related to $\Phi$. When biased with constant $I$, periodic modulation of $V(\Phi)$ is shown in Fig. 2a. The sensitivity of the SQUID becomes maximum, $|\partial V/\partial \Phi| \sim 39$ μV/$\Phi_0$, near $I_c$ (see Fig. 2b), which is similar value to previous result in InAs NW SQUID.[19] Another plot of differential

resistance, $dV/dI$, with $\Phi$ exhibits similar behavior at low bias current. At high bias above $I = 0.8$ μA, however, π-phase-shifted oscillations occur in Fig. 2c. This π-junction behavior in the weak link has been attributed to the existence of ferromagnetic layer,[28] nonequilibrium electron distribution in normal region,[29] quasiparticle-pair interference effect[30] or a quantum dot[17] between two superconducting electrodes. Numerical differentiation of the $I$-$V$ curves, however, reveals that the π-phase shift observed in this work is a direct result of the flux-dependent $dV/dI$ vs. $I$ curves, as shown in Fig. 2d. Near zero bias current, $dV/dI$ becomes maximal at half-integral flux quantum and minimal at integral one. At higher bias current above $I = 0.8$ μA, the opposite behavior is obtained to induce the π-phase shift. In the intermediate bias region, $h/4e$ oscillations instead of $h/2e$ ones are observed. A metallic dc SQUID made of Al-Au-Al junctions exhibits similar phenomenon.[31]

Color plot of $dV/dI$, obtained from device D2 at much lower $T$, is displayed in Fig. 3a-d. It is noted that there are several differences between Fig. 3a-d and Fig. 1c. Firstly, the supercurrent "off" state ($I_c = 0$) is not seen in D2, but the periodic modulation of $I_c(\Phi)$ between the maximum ($I_{c,max}$) and minimum ($I_{c,min}$) values of $I_c$. Since the screening effect is negligible ($\beta_L \ll 1$), the absence of the $I_c$-off state can be caused by the asymmetric weak links,[14] where $I_{c1}$ and $I_{c2}$ are estimated by $I_{c1} = (I_{c,max} + I_{c,min})/2$ and $I_{c2} = (I_{c,max} - I_{c,min})/2$. This results in $I_{c1} = 510$ nA, $I_{c2} = 150$ nA and $I_{c2}/I_{c1} = 0.29$ for D2 at $T = 0.3$ K. The asymmetric $I_c$'s are also responsible for the shift of $I_c(\Phi)$ curves in opposite polarity.[14] Secondly, the skewed $I_c(\Phi)$ curves are obtained instead of the sinusoidal ones. The skewness can be defined by $(2\varphi_{max}/\pi - 1)$, where $\varphi_{max}$ is the position of $I_{c,max}$.[32] Figure 3e shows that the skewness decreases monotonously with temperature, evolving into a sinusoidal $I_c(\Phi)$ curve at higher temperature, as observed in D1. The skewness can be caused by the asymmetric $I_c$'s in two weak links in the SQUID. The ratio $I_{c2}/I_{c1}$, however, decreases with temperature to enhance the $I_c$ asymmetry, resulting in $I_{c2}/I_{c1} = 0.17$ at $T = 2.7$ K (see Fig. S2), which is contrary to the temperature dependence of the skewness.

A more plausible explanation can be found in the CPR of the NW-based superconducting weak link. It is well known that the CPR in the superconducting weak links is nonsinusoidal at low temperatures and converts into a sinusoidal one near $T_c$.[33] For a short and diffusive weak link, the CPR is given by

$$I_c(\varphi) = \frac{4\pi k_B T}{eR_N} \sum_{\omega>0} \frac{\Delta \cos\left(\frac{\varphi}{2}\right)}{\delta} \arctan \frac{\Delta \sin\left(\frac{\varphi}{2}\right)}{\delta} \qquad (2),$$

where $R_N$ is the normal-state resistance of the junction, $\delta = \sqrt{\Delta^2 \cos^2(\varphi) + (\hbar\omega)^2}$, $\hbar\omega = \pi k_B T(2n + 1)$ is the Matsubara energy, $\Delta$ is the superconducting energy gap, $\varphi$ is the phase difference between two superconducting electrodes, and $n$ is an integer.[33] After applying the Eq.2 for $I_{c1}$ and $I_{c2}$ in combination with flux quantization in the SQUID loop, the calculation results are depicted in Fig. 3f as solid lines, which are in good agreement with the $I_c(\Phi)$ data. $\Delta(T = 0) = 1.04$ meV was used as a fitting parameter, which is consistent with the experimental value (see Fig. S3). Since the elastic mean free path and the Thouless energy of the PbS NW are obtained to be $l_e = 26$ nm and $E_{Th} = \hbar D/L^2 = 112$ μeV, respectively, where $D = 103$ cm$^2$/s is the diffusion coefficient and $L = 250$ nm is the length of the superconducting weak link for D1, the PbS NW weak link is in a short ($E_{Th}/\Delta \sim 0.11$) and diffusive ($l_e \ll L$) junction regimes.

Figure 4a-d display color plots of $dV/dI$ as a function of magnetic flux and bias current at different gate voltages, $V_g$. Note that the $V_g$-dependent change of $I_c(\Phi)$ curves is quite similar to the temperature-dependent one shown in Fig. 3a-d. As the gate voltage is decreased, $I_c$ also decreases and the skewness becomes smaller, as shown in Fig. 4e. We used Eq.2 for fitting the short and diffusive junction model to the $V_g$-dependent $I_c(\Phi)$ data, where the effective temperature, $T_{eff}$, was used as a fitting parameter. The fitting results (solid lines in Fig. 4f) are in good agreement with the $I_c(\Phi)$ data (symbols), while yielding $T_{eff}$ in a reasonable range (see Fig. 4e). With the application of negative $V_g$, $I_c$ is suppressed and the distorted CPR becomes sinusoidal. These features are quite similar to the effects of increased temperature, as shown in Fig. 3, and thus $T_{eff}$ increases at negative $V_g$. To the best of our knowledge, this work is the first to demonstrate the gate-voltage dependence of the CPR in the NW-based SQUIDs. Since the CPR determines the shape of an anharmonic potential well for the Josephson phase particle,[23] our observed gate-tunable CPR would be important for developing nanohybrid superconducting qubits.

**Acknowledgment**

This work was supported by the U.S. National Science Foundation (Grant DMR-1310678 for DY) and the National Research Foundation of Korea through the Basic Science Research Program (Grant No. 2015R1A2A2A01006833 for YJD).

**Figure Captions**

**Figure 1.** (a) False-colored SEM image of a typical NW SQUID with a PbS NW (yellow) and PbIn superconducting electrodes (blue). Yellow dashed line indicates an effective area of the supercurrent loop. Bias current flows from I+ to I−, while voltage is measured between V+ and V−. (b) Current-voltage curves at $T = 2.8$ K with different magnetic flux through the NW SQUID loop (sample D1). Inset: Temperature dependence of the critical current. Line is a guide to the eye. (c) Color plot of differential resistance, $dV/dI$, as a function of magnetic flux and dc bias current. The black-colored region is the supercurrent regime and the white

line indicates a sinusoidal current-phase relation described by Eq.(1). (d) Temperature-dependence of $I_c(\Phi)$ relation. Line is a fit using Eq.(1).

**Figure 2.** (a) Modulation of output voltage as a function of $\Phi$ at $T = 2.7$ K. Bias current, $I$, increases from 0.1 to 1.0 µA in steps of 0.1 µA from bottom to top. (b) Flux-to-voltage transfer function with different $I$. Error bars are indicated. (c) $dV/dI$ vs. $\Phi$ curves with $I = 0$, 0.4, 0.5, 0.6, 0.7, 0.8, and 1.0 µA, respectively, from bottom to top. The lock-in bias current is $I_{ac} = 200$ nA. (d) $dV/dI$ vs. $I$ curves with different $\Phi$.

**Figure 3.** Color plot of $dV/dI$ as a function of $I$ and $\Phi$ at (a) $T = 0.3$, (b) 1.5, (c) 2.1, and (d) 2.7 K for sample D2 with $V_g = 45$ V. The white lines indicate $I_c(\Phi)$ curves at different temperatures, respectively. (e) Temperature dependence of skewness. (f) $I_c(\Phi)$ data (symbols) with different $T = 0.3$, 0.9, 1.5, 2.1, and 2.7 K, respectively, from top to bottom. The solid lines are fitting results using Eq.2 in the text.

**Figure 4.** Color plot of $dV/dI$ as a function of $I$ and $\Phi$ at (a) $V_g = 45$, (b) 20, (c) 0, and (d) -60 V at $T = 0.3$ K. The white lines indicate $I_c(\Phi)$ curves at each gate voltage. (e) Skewness and $T_{eff}$ as a function of $V_g$ at $T = 0.3$ K. The lines are guides to the eye. (f) $I_c(\Phi)$ data (symbols) with different $V_g = 45$, 20, 0, -30, and -60 V from top to bottom. The solid lines are fitting results.

**Figure 1**

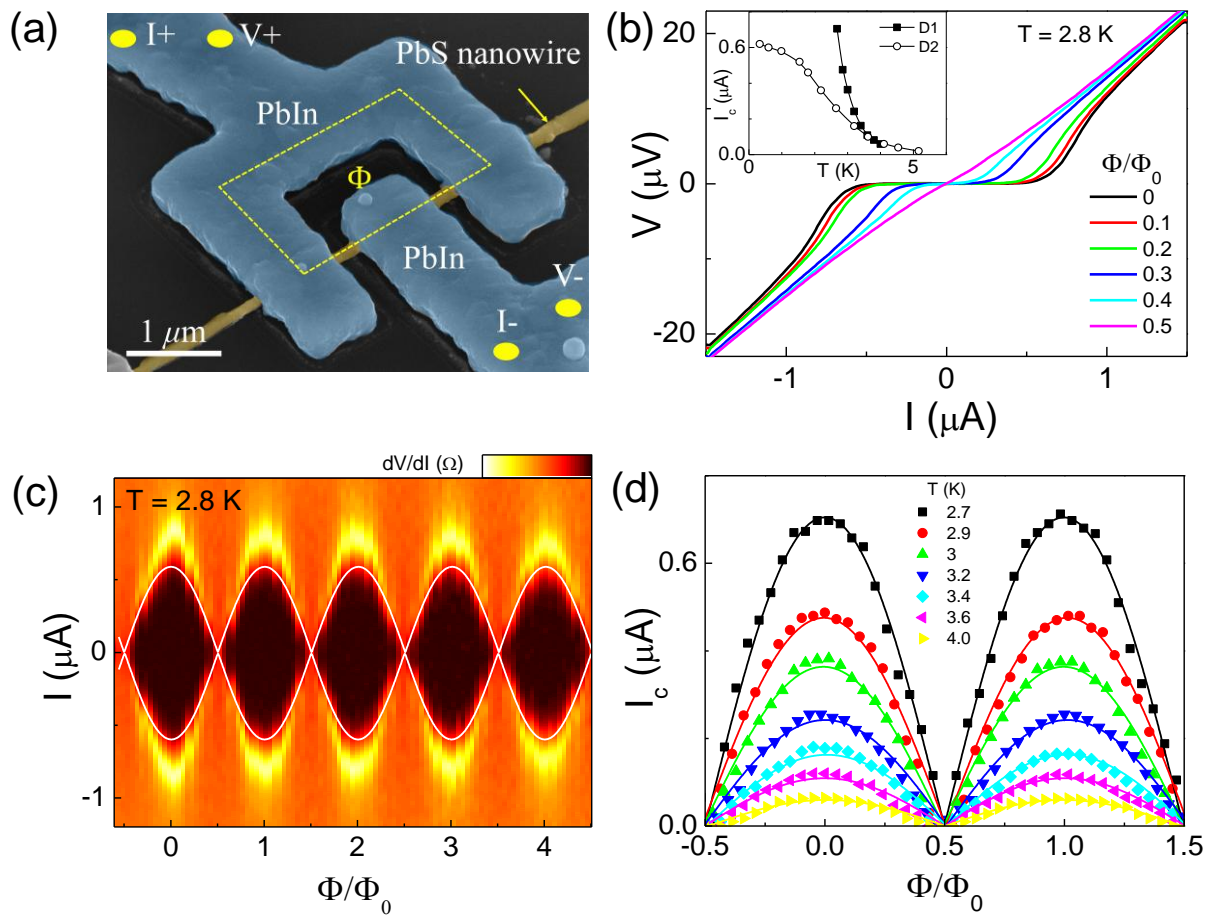

**Figure 2**

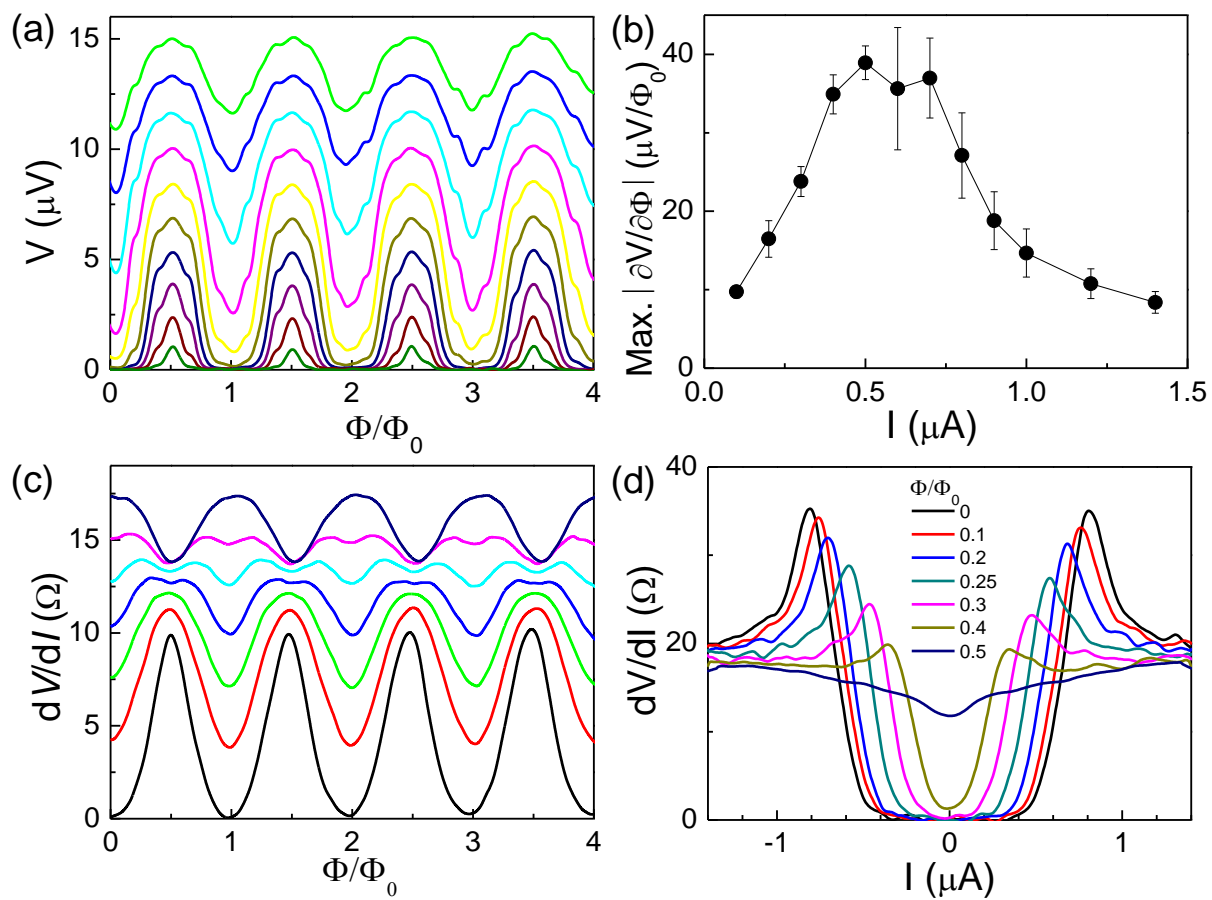

**Figure 3**

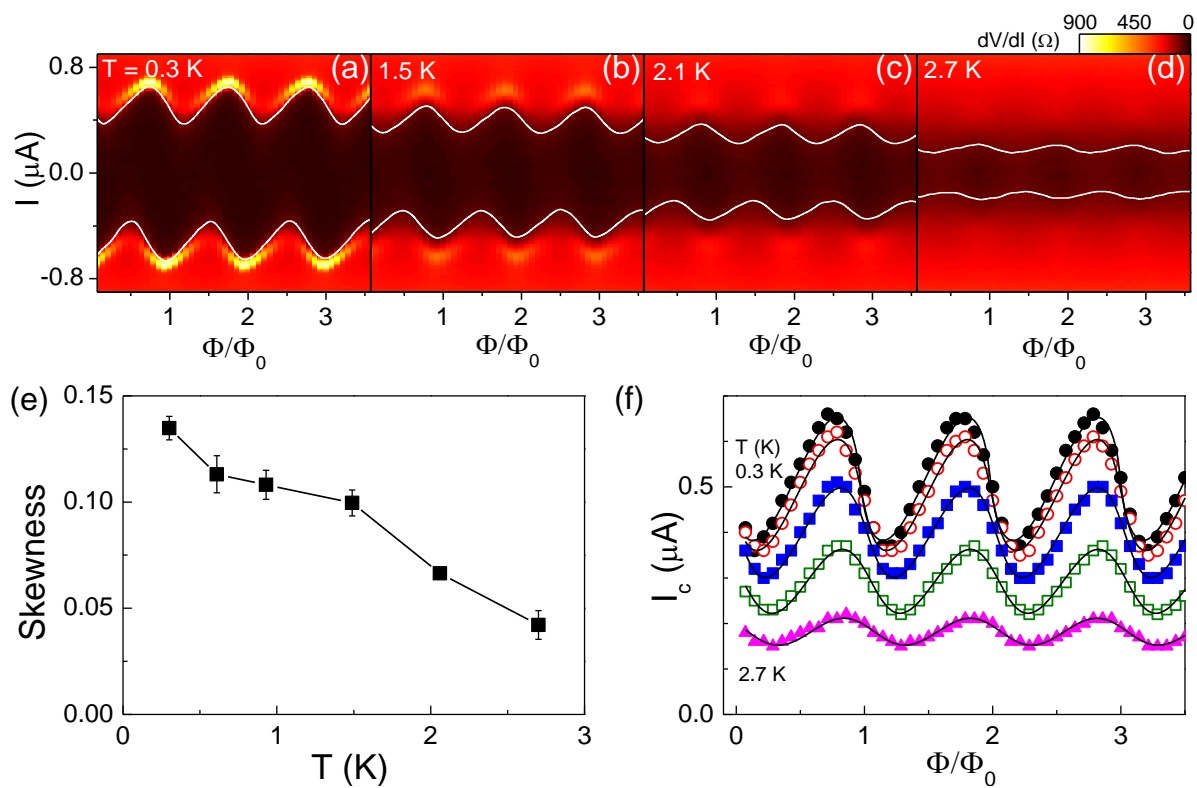

**Figure 4**

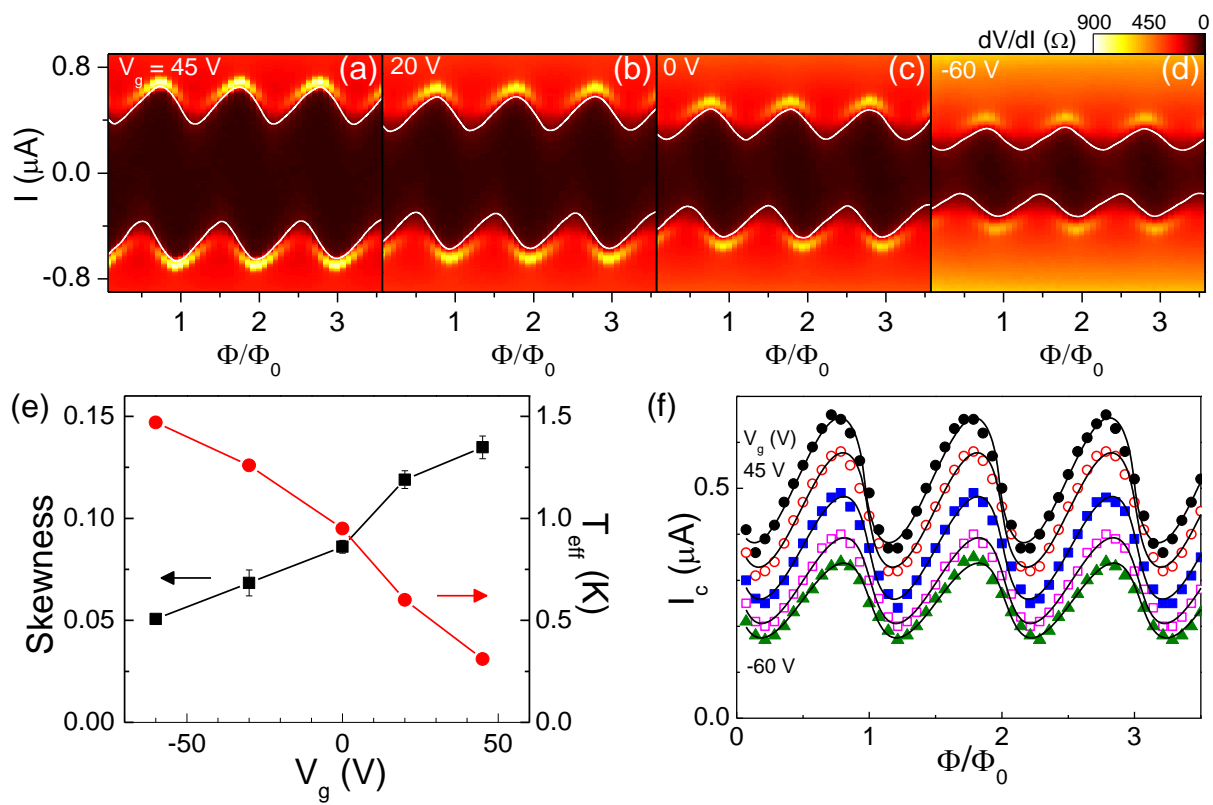